\documentclass[aps,prd,groupedaddress,nofootinbib,12pt]{revtex4-2}
\usepackage{comment}
\usepackage{ulem}
\usepackage{dsfont}
\usepackage{natbib}
\usepackage{amssymb}
\usepackage{graphicx}
\usepackage{amsmath}
\usepackage{url}
\usepackage{bm}
\usepackage{epstopdf}
\usepackage{color}
\usepackage[dvipsnames,svgnames]{xcolor}
\usepackage{comment}
\usepackage{enumerate}
\usepackage{enumitem}
\usepackage{mathrsfs}
\usepackage[colorlinks=true,breaklinks]{hyperref}
\newcommand{\gaborTwo}{\textcolor{blue}}

\newcommand{\jack}{\textcolor{Green}}
\newcommand{\di}{\partial}
\newcommand{\f}{\mathscr{F}}

\def\be {\begin{equation}}
	
	\def\ee  {\end{equation}}

\def\bea {\begin{eqnarray}}
	
	\def\eea {\end{eqnarray}}

\newcommand{\nn} {\nonumber}

\def\half {\frac{1}{2}}

\begin{document}

	\title{Spontaneous Symmetry Breaking To GR in SO(4,2)  Gravitational Yang-Mills Theory }
	
	\author{Jack Gegenberg}
	\email{geg@unb.ca}
	\affiliation{Department of Mathematics and Statistics, University of New Brunswick, Fredericton, NB, Canada E3B 5A3}
	\author{Gabor Kunstatter}
	\email{g.kunstatter@uwinnipeg.ca}
	\affiliation{Department of Physics, University of Winnipeg, Winnipeg, MB Canada R3B 2E9}

	\date{\today}
	
	\begin{abstract}
		
		We consider a Yang-Mills type gauge theory of gravity based on the conformal group SO(4,2) coupled to a conformally invariant real scalar field. The goal is to generate fundamental dimensional constants via spontaneous breakdown of the conformal symmetry. In the absence of other matter couplings the resulting theory resembles Weyl-Einstein gravity, {with the Newton constant given by the square of the (constant) vacuum expectation value of the scalar,  the cosmological constant determined by the quartic coupling constant of the scalar field and the Weyl to Einstein transition scale determined by the Yang-Mills coupling constant.} The emergent theory in the long-wave-length limit is Einstein gravity with cosmological constant.  As an illustrative example we present an exact spherically symmetric cosmological solution  with perfect fluid energy-momentum tensor that reduces to $\Lambda$FRW in the long-wavelength limit.

	\end{abstract}
	
	\maketitle
	\clearpage
	\clearpage

	\section{Introduction}
	
	General relativity (GR) is a well-tested theory of gravity on scales ranging from microns to the size of the solar system.  However, efforts to extend GR outside of this range have not succeeded. {Near the Planck scale, a quantum description of  gravity is generally deemed to be necessary  but has so far has proved illusive. At very large scales} cracks  appear in the edifice of observational support for the theory.  The gravitational interaction over galactic distances requires dark matter to account for rotation curves and dynamics of clusters, while over cosmological distances we require dark energy to drive accelerated expansion.  While it is quite possible, and some may argue probable, that one or both of these effects have an explanation within particle physics, it seems worthwhile to explore the possibility that GR itself is not the correct theory of gravity at both cosmic and Planck scales.  Rather, perhaps gravity is emergent at scales between the micron and the size of the solar system.  The model under discussion here explores an attempt at such a formulation.
	
	There is a large and growing literature on the construction of emergent gravity, beginning perhaps with the work of Sakharov \cite{Sakharov} in the 1960s, and carried forward by many others, including S.L. Adler \cite{Adler}, T. Padmanabhan, \cite{Padmanabhan}, E. Verlinde \cite{Verlinde} and others.  The approach taken in this stream of research begins with matter interacting via the gauge forces of the Standard Model, but with no gravitational interaction.  The latter emerges from quantum effects.

	The approach taken here is to replace GR with a theory respecting local conformal (Weyl) symmetry.  There are diverse motivations for such an approach (see \cite{Hooft:2014daa} for a recent discussion). One of the earliest involves {the observation that Maxwell's equations are invariant not only under the Poincare group of transformations ISO(3,1), but also under the larger SO(4,2) group of conformal transformations of Minkowski spacetime \cite{cunningham,*bateman}. This symmetry is shared by other important massless field theories \cite{dirac,*drew}, including the standard model with zero Higgs mass.}  It seems logical that the theory of gravity should have the same symmetries as electromagnetism, at least locally, yet GR retains local ISO(3,1), not SO(4,2), symmetry.

	Shortly after the appearance of GR, Weyl \cite{weyl} and Bach \cite{bach} tried to rectify this by writing down a locally conformally invariant theory of gravity  and electromagnetism.  This approach failed as a unified theory because it gave unacceptable gravity-EM couplings and could not reproduce solar system dynamics as linear gravity was governed by a fourth order Poisson equation.  To recover the inverse square law in conformal gravity, Mannheim has recently suggested that point sources are described by highly singular distributions involving derivatives of $\delta$-functions \cite{Mannheim:2007ug,Mannheim:1992tr,*Mannheim:1999bu,*Mannheim:2011ds}.

{A second important motivation for considering conformally invariant theories  is that Einstein gravity is not perturbatively renormalizable due primarily to the fact that Newton's constant has non-trivial dimension ($L^{-2}$). In an effort to circumvent this problem, conformal gravity replaces the Einstein-Hilbert first order action by terms quadratic in the  {Levi-Civita, that is, torsion-free} curvature tensor. Such an action has a dimensionless coupling constant but the price paid is that the equations of motion are {\it quartic} in time derivatives of the metric tensor. {Such theories are thought to be generically plagued by ghosts, although counter examples do exist \cite{Bender:2008prl},\cite{Bender:2008gh}}
}

   Here we  consider an alternative approach that has a long history, namely theories of gravity  based on the Yang-Mills (YM) action.  In 4D curved space time the YM action  
for compact gauge group is uniquely (up to topological terms) given by:
	 \be
	S= - \frac{1}{2 g_\text{YM}^{2}}\int d^4x \sqrt{-g}g^{\mu\nu}g^{\alpha\beta}\text{Tr}( \textbf{F}_{\mu\alpha} \textbf{F}_{\nu\beta}),
	\label{eq:YMaction}
	\ee
{}{is quadratic in the curvature $\mathbf{F}_{\alpha\beta}$ of a Lie-algebra valued connection $\textbf{A}_{\alpha}$ on a principle bundle over the spacetime manifold $(M,g)$, where $g_{\alpha\beta}$ is a {\it given} non-dynamical Lorentzian metric on $M$. If the gauge group is a `spacetime symmetry group', e.g. the Poincare, deSitter, anti-deSitter or the 15 parameter conformal group, then the action has two metrics.  Besides the fixed background, there is a second metric available, one that is dynamical. It is determined by the translation component of the gauge potential.  
}

{}{		
	The model we discuss below dispenses with the fixed background metric and considers only the  dynamical metric determined by the translation component of the gauge potential. It is this metric that is used to define the integration measure and contract the space time indices of the Yang-Mills curvature tensors in (\ref{eq:YMaction}). The coupling constant $g_\text{YM}$ is still dimensionless, allowing for the possibility that the theory is perturbatively renormalizable.  
	Moreover, the equations of motion are at most second-order in time derivatives of the gauge potential  unless we impose conditions which relate the spin-connection to the metric.  Specifically,  in the torsion free sector they reduce to equations that, like conformal gravity, contain terms that are higher order derivatives of the metric. The hope  nonetheless is that 
when the full spectrum of fluctuations is taken into account, including torsion, 
the quantum theory based on (\ref{eq:YMaction}) may  avoid the ghost problems associated with higher order theories. 
}

	{The present work extends and deepens early work by one of us:\cite{Gegenberg_2016,Gegenberg_2017,gegenberg2018gravitational}.  In those papers, the formalism of a gravitational theory based on an SO(4,2) Yang-Mills type action was developed, and it was shown that it is consistent with linearized GR {and is able to} describe a cosmology that includes inflation and accelerated expansion. {The vacuum sector of the theory studied in \cite{Gegenberg_2016,Gegenberg_2017,gegenberg2018gravitational} does not contain GR in its low energy limit.} Here we address this limitation }by adding a matter action with a conformal symmetry that is spontaneously broken in the vacuum, giving a theory that  includes GR.  The theory differs from the original Weyl-Bach theory
		in that it is {\it a priori} quadratic in time derivatives of the fields.
	
	To construct a YM gauge theory of gravity with action (\ref{eq:YMaction}) one first chooses a gauge group, and then identifies {a subset of }the components of the gauge potential $\mathbf{A}_{\alpha}$ with a vierbein $e^{a}_{\alpha}$ (or soldering form) and spin-connection $\omega^{ab}_{\alpha}$ on $(M,g)$.  This introduces an explicit algebraic dependence of the metric on the gauge potential, implying that YM gravity models generally exhibit less gauge symmetry than their purely YM cousins.  \jack{U}nlike the Einstein-Cartan action, the torsion of the connection is not constrained to be zero \cite{Utiyama:1956sy, kibble,*townsend,*macman,*PhysRevLett.42.1021,*ivnied,Wheeler:1991ff,*Hazboun:2013lra,*Wheeler:2013ora,Huang:2008ik,*Huang:2009nc,*Huang:2009sk,*Lu:2013sai,*Lu:2013bt}.
	
	If one selects one of the Poincare, de Sitter, or anti-de Sitter groups as the gauge group, the identification of  $e^{a}_{\alpha}$ and spin-connection $\omega^{ab}_{\alpha}$ exhausts all the components of $\mathbf{A}_{\alpha}$.  (The de Sitter case has been investigated in \cite{Huang:2008ik,*Huang:2009nc,*Huang:2009sk,*Lu:2013sai,*Lu:2013bt}.)  Here, we follow Wheeler {\it et al } \cite{Wheeler:1991ff,*Hazboun:2013lra,*Wheeler:2013ora} and consider the gauging of the SO(4,2) group.  In this case, there are extra components of the gauge potential associated with special conformal transformations and dilatations, allowing for more general matter\jack{-}gravity coupling than in GR.  The resulting equations exhibit local conformal symmetry.\\

	The paper is organized as follows. In Section \ref{sec:TheModel} we specify the lie algebra of the YM action, write  the gauge potentials in terms of geometrical variables and specify the  gauge transformations of the geometric variables. We show why the theory is invariant under only { infinitesimal gauge transformations given by a sub-algebra of the full YM gauge algebra.} Of particular interest is the generator of local scale, or conformal,  transformations. Section \ref{sec:ConformalScalar} adds a conformally coupled scalar field to the action, and shows that the resulting theory is gauge equivalent to Einstein-Weyl gravity with {}{ Newton's constant and cosmological constant emerging via the breaking of the conformal symmetry.} In Section \ref{eq:CosmologicalSolution} we further add a non-isotropic perfect fluid energy momentum tensor, and derive a family of exact spherically symmetric, inhomogeneous cosmological solutions to the full set of equations in the model. The solutions reduces to the $\Lambda$FRW solution with radiation at large distances. We close in Section \ref{sec:Discussion} with a summary and prospects for future work. {}{There are two appendices.  Appendix \ref{sec:Geometry}, presents the Bach tensor and Einstein tensor for our metric ansatz and solutions while \ref{sec:GeneralRho} provides our cosmological solution in a general class of conformal gauges.}
	
	\section{Action, Gauge Invariance and Equations of Motion}
	
	\label{sec:TheModel}

	 We base our model on the YM action (\ref{eq:YMaction}) with gauge group, SO(4,2) which is the conformal group of Minkowski spacetime. For details, see \cite{Gegenberg_2016}. The spacetime components of the gauge potentials are given by 
	\begin{equation}
		\mathbf{A}_{\alpha} = A^{A}_{\alpha} \mathbf{J}_{A} =   {m}e^{a}_{\alpha} \mathbf{P}_{a} +\frac{1}{m}\ell^{a}_{\alpha} \mathbf{K}_{a} + \omega_{\alpha}^{ab} \mathbf{J}_{ab} + q_{\alpha} \mathbf{D},
	\end{equation}
	where the upper-case latin indices $A, B,...$ denote the 15 dimensions of the Lie algebra of SO(4,2) and  $\left\{\mathbf{P}_{a},\mathbf{K}_{a},\mathbf{J_{ab}}_{a},\mathbf{D}\right\}$ are the 15 generators of SO(4,2).  We break SO(4,2) gauge invariance in the Yang-Mill action by identifying the metric tensor $g_{\mu\nu}(x):=\eta_{ab} e^a_\mu(x) e^b_\nu(x)$ with the gauge potential $e^a_\mu$.  The factor $m$, which has dimensions of inverse length, is inserted for convenience, so that the metric tensor is dimensionless. 
The quantities $l^a_\mu$  as defined above have dimensions of length${}^{-2}$, which is appropriate since  $f_{\mu\nu}=\eta_{ab}e^a_\mu l^b_\nu$ will in effect play the role of a field strength in the gravitational theory. The arbitrary parameter $m$ will disappear from the gravitational action as required by the conformal invariance of the action.

	The components of the field strength associated with the potentials are denoted by $\mathbf{F}_{\alpha\beta} = F^{A}_{\alpha\beta} \mathbf{J}_{A}$ and are given by
	\begin{equation}
		F^{A}_{\alpha\beta} = \di_{\alpha} A^{A}_{\beta} - \di_{\beta} A^{A}_{\alpha} + f^{A}{}_{BC} A^{B}_{\alpha} A^{C}_{\beta},
	\end{equation}
	where  the structure constants are denoted by $f^{C}{}_{AB} \mathbf{J}_{C}$.
	
	{}{Our goal is to write down a diffeomorphism invariant action without introducing additional structure beyond what is provided by the so(4,2) gauge algebra. To this end, we} identify various components of $\mathbf{A}_{\alpha}$ in the $\mathbf{J}_{A}$ basis with geometric quantities in a 4-dimensional Lorentzian manifold $M$ with metric $g_{\alpha\beta}$ and affine connection $\Gamma^{\alpha}{}_{\beta\delta}$.  In particular, we take $e^{a}_{\alpha}$ as the components of an  orthonormal frame field on $M$, with $\omega^{ab}_{\alpha}$ as the associated connection one-forms.  Hence, the metric and connection are given by:
	\begin{equation}\label{eq:tetrad postulate}
		g_{\alpha\beta} = \eta_{ab} e^{a}_{\alpha} e^{b}_{\beta}, \quad \Gamma^{\gamma}{}_{\alpha\beta} = e^{\gamma}_{a}(\di_{\alpha} e_{\beta}^{a}  + \omega^{ac}_{\alpha} e_{c\beta}).
	\end{equation}
	In these expressions, lowercase Greek and Latin indices are raised and lowered with $g_{\alpha\beta}$ and $\eta_{ab}$, respectively.  The curvature one-forms are anti-symmetric in their frame indices $\omega^{(ab)}_{\alpha} = 0$, from which it follows that the affine connection is metric compatible \cite{Carroll:2004st}:
	\begin{equation}
		0 = \nabla_{\alpha} g_{\beta\gamma},
	\end{equation}
	where $\nabla_{\alpha}$ is the derivative operator defined by $\Gamma^{\alpha}{}_{\beta\delta}$.  The Riemann curvature and torsion tensors of $M$ are given by:
	\begin{align}
		R^{\mu\nu}{}_{\alpha\beta} & = e^{\mu}_{a} e^{\nu}_{b} (d\omega^{ab} + \omega^{ac} \wedge \omega_{c}{}^{b})_{\alpha\beta}, \\
		T^{\alpha}{}_{\beta\gamma} & = e^{\alpha}_{a} (de^{a} + \omega^{ac}\wedge e_{c})_{\beta\gamma}.\label{eq:torsion def}
	\end{align}
	Note that we do {\it not} assume $T^{\alpha}{}_{\beta\gamma} = 2\Gamma^{\alpha}{}_{[\beta\gamma]}=0$.
	
	We now re-write the  Yang-Mills action (\ref{eq:YMaction}) choosing  specific coordinate and lie algebra bases and adding a generic matter action for future reference:
	\begin{align}
		S = - \frac{1}{2g^{2}_\text{YM}} \int d^{4}x \sqrt{-g} g^{\alpha\mu} g^{\beta\nu} h_{AB} F^{A}_{\alpha\beta} F^{B}_{\mu\nu} + S_\text{m},\label{eq:action}
	\end{align}
	Here, $S_\text{m} = S_\text{m}[A_{\alpha}^{A},\psi]$ is the action for a matter field $\psi$ and 
	the non-trivial components of the Cartan-Killing metric $h_{AB}$ are:
	\begin{gather}
		\nn h_{a\bar{b}} = h_{\bar{a}b}=-2\eta_{ab}, \quad h_{14,14} = 2.\\
		h_{[ab][cd]} = h_{[cd][ab]}=-4\eta_{a[c}\eta_{d]b}.
		\label{eq:KillingMetricComponents}
	\end{gather}
	The notation here is that $a,\bar{a}=0,1,2,3$ denote components in the direction of translations $\mathbf{P}_a$ and special conformal transformations $\mathbf{K}_a$, respectively.  The six indices $[ab]$ consist of $[12],[23],[31],[01],[02],[03]$ and denote directions along  the distinct non-zero generators $\mathbf{J}_{ab}$ of Lorentz transformations.  Finally, the index {$[14]$} denotes the component in the direction of the generator $\mathbf{D}$ of dilatations.
	
	Eqs.~(\ref{eq:action}) and  (\ref{eq:KillingMetricComponents}) are the defining relationships for our model.  
	

		Making use of the definitions above, we can rewrite the action (\ref{eq:action}) exclusively in terms of spacetime tensors:
		\begin{multline}\label{eq:spacetime action}
			S = \frac{2}{g_\text{YM}^{2}} \int d^{4}x \sqrt{-g} \Big[ (R_{\alpha\beta\gamma\delta} - {\frac{1}{2}} \phi_{\alpha\beta\gamma\delta})^{2} + 2 (\nabla^{\alpha}f^{\mu\beta}+ f^{\mu\sigma}T_{\sigma}{}^{\alpha\beta}-f^{\mu\alpha}q^{\beta}) \times  (T_{\mu\alpha\beta} + g_{\mu[\alpha} q_{\beta]} ) + \\ (\di_{[\alpha} q_{\beta]} + \f_{\alpha\beta})^{2}  \Big] + S_{m}.
		\end{multline}
		Here, we have defined:
		\begin{gather}
			\nn f_{\alpha\beta} := \eta_{ab} e_{\alpha}^{a} l_{\beta}^{b}, \quad \f_{\alpha\beta} := {\frac{1}{2}} f_{[\alpha\beta]}, \\
			\phi^{\alpha\beta\gamma\delta}  := g^{\gamma[\alpha} f^{\beta]\delta}  - g^{\delta[\alpha} f^{\beta]\gamma}. \label{eq:phi def}
		\end{gather}
		As anticipated, the arbitrary dimensional  parameter $m$ does not appear in the above action.
		It was shown in \cite{Gegenberg_2016} that under certain restrictive circumstances, (\ref{eq:spacetime action}) reduces to the action of Weyl-squared gravity.

		Variation of the action (\ref{eq:action}) with respect to the gauge potential yields the equation of motion:
		\begin{equation}\label{eq:EOM 1}
			D_{\mu} F^{B\mu\nu} = k^{B\nu} + j^{B\nu},
		\end{equation}
		where $D_{\mu}$ is the gauge covariant derivative:
		\begin{equation}
			D_{\mu} F^{B\mu\nu} := \hat\nabla_{\mu}  F^{B\mu\nu} + f^{B}{}_{CD} A^{C}_{\mu} F^{D\mu\nu},
		\end{equation}
		and $\hat\nabla_{\mu}$ is the derivative operator defined from the Levi-Civita connection:
		\begin{equation}
			\hat{\Gamma}^{\alpha}{}_{\beta\gamma} = \tfrac{1}{2} g^{\alpha\rho} (\di_{\beta} g_{\rho\gamma} + \di_{\gamma} g_{\rho\beta} - \di_{\rho} g_{\beta\gamma}).
		\end{equation}
		Note that $\nabla_{\alpha} = \hat\nabla_{\alpha}$ if and only if $T^{\alpha}{}_{\beta\gamma}=0$.  The currents in (\ref{eq:EOM 1}) are given by:
		\bea
		k^{A\nu} &=& \frac{\tau^{\mu\nu} e^{b}_{\mu} \mathbf{P}_{b}}{2}
		\label{eq:Currentk}\\
		j^{B\,\nu} &=& - \frac{g_\text{YM}^{2}}{2\sqrt{-g}} h_{BC}\frac{\delta(\sqrt{-g} \mathcal{L}_\text{m})}{\delta A^{C}		_{\nu}},
		\label{eq:Currentj}
		\eea
		where $j^{B\,\nu} $ is the current associated with the matter Lagrangian density, $\mathcal{L}_\text{m}$. The quantity $k^{B\,\nu}$ is the current obtained from varying the Yang-Mills action with respect to $e^a_\mu$ and is proportional to the Yang-Mills stress energy tensor: 
		\begin{equation}
			\tau^{\rho\sigma} := h_{AB}\left( F^{A\rho\mu} F^{B\sigma}{}_{\mu} - \tfrac{1}{4} g^{\rho\sigma} F^{A}_{\mu\nu} F^{B\mu\nu} \right).
		\end{equation}
		
		
		Due to the  explicit dependence of the metric on the components  of the potential along the translation generators, the YM-like action (\ref{eq:action}) is \emph{not} invariant under the full algebra of usual infinitesimal YM gauge transformations, even with no `external' matter present.  The residual gauge symmetry is eleven dimensional, with infinitesimal gauge transformation parametrized by \cite{Gegenberg_2016}: 
		\begin{equation}\label{eq:gauge 1}
			\epsilon^{A} \mathbf{J}_{A} =\lambda^{a} \mathbf{K}_{a} + \Lambda^{ab} \mathbf{J}_{ab} + \Omega \mathbf{D}.
		\end{equation}
		{}{where all $\epsilon^A$ are dimensionless.}
		The absence of the $\mathbf{P}_{a}$ generators in (\ref{eq:gauge 1}) is not surprising: as in GR, the Poincar\'{e} translational symmetry of the Minkowski metric is supplanted by the diffeomorphism invariance of curved space in our model.\footnote{When thinking about gauge transformations in this model, one may be tempted to interpret $\mathbf{P}_{a}$ as the generator of \emph{diffeomorphisms} rather than \emph{translations}.  
			The action (\ref{eq:action}) is invariant under diffeomorphisms but not invariant under gauge transformations generated by $\mathbf{P}_{a}$.  Therefore, $\mathbf{P}_{a}$ is not the generator of diffeomorphisms.
		}
		
		{}{Under  gauge transformations of the form (\ref{eq:gauge 1}) ,} the frame fields and metric transform as
		\begin{equation}\label{eq:gauge 3}
			\delta e^{a}_{\alpha} = (-\Lambda^{a}{}_{b} + \tfrac{1}{2} \Omega \delta^{a}{}_{b}) e^{b}_{\alpha}, \quad \delta g_{\alpha\beta} = \Omega g_{\alpha\beta}.
		\end{equation}
		We see that $\Lambda^{ab}$ generates infinitesimal Lorentz rotations of the frame fields and $\Omega$ generates infinitesimal local conformal transformations.  Hence, the invariance of the action under (\ref{eq:gauge 3}) means that our model is conformally invariant.
		
		Under the transformation (\ref{eq:gauge 1} ) the component of $\mathbf{A}_{\alpha}$ parallel to $\mathbf{D}$ transforms as
		\begin{equation}
			\delta q_{\alpha} = \di_{\alpha}\Omega + {\frac{1}{2}} \lambda_{\alpha}.
		\end{equation}
		{}{It is obvious that the field $q_\alpha$ is pure gauge. That is, }via a simple series of gauge transformations of the form $\epsilon^{A} \mathbf{J}_{A} = \lambda^{a} \mathbf{K}_{a}$ we can set  $q_{\alpha}$ to anything we want, {including zero}.
		In the {following}
		we will restrict to the set of gauge transformations :
		\begin{equation}\label{eq:gauge 2}
			\epsilon^{A} \mathbf{J}_{A} = -2e^{a\alpha} \di_{\alpha} \Omega \, \mathbf{K}_{a} + \Lambda^{ab} \mathbf{J}_{ab} + \Omega \mathbf{D}.
		\end{equation}
		We refer to this seven parameter algebra 
		of transformations as $g_7$.
		
		The gauge transformation (\ref{eq:gauge 2}) generates a conformal transformation of the metric that not only preserves the gauge condition $q_\alpha = 0$, but is  also linear in the torsion.  
		Specifically, under this transformation, the torsion tensor transforms as
		\be
		\delta T^a_{\alpha\beta} = \tfrac{1}{2} T^a_{\alpha\beta} \Omega -\Lambda^a_{{\,\, }b} T^b_{\alpha\beta}
		\ee
		Thus, under transformations that preserve $\delta q_\alpha$, the torsion tensor transforms homogeneously and the torsion-free sector is preserved under this restricted group of gauge transformation.

		
		
		We rewrite the matter current as
		\begin{equation}
			j^{B\nu} \mathbf{J}_{B} =  a^{a\nu} \mathbf{P}_{a} + b^{a\nu} \mathbf{K}_{a} + c^{ab\nu} \mathbf{J}_{ab} + d^{\nu} \mathbf{D},\label{eq:matter currents}
		\end{equation}
		and define $a^{\alpha\nu} := e^{\alpha}_{a} a^{a\nu}$, etc.  The structure of $h_{AB}$ implies that $a_{\alpha\beta}$ is proportional to the functional derivative of the matter action with respect to $l^{a}_{\alpha}$; that is, $a_{\alpha\beta}$ characterizes matter which couples to the generators of special conformal transformations.  Similarly, $b_{\alpha\beta}$, $c_{\alpha\beta\gamma}$, and $d_{\alpha}$ describe matter with coupling to $e^{a}_{\alpha}$, $\omega^{ab}_{\alpha}$, or $q_{\alpha}$, respectively.

		The equations of motion, expressed in terms of spacetime tensors, are:
		\begin{subequations}
			\begin{eqnarray}
				\nonumber a^{\alpha\nu}&=&  ({\hat \nabla}_{\mu}-T^{\sigma}{}_{\sigma\mu}-\tfrac{1}{2} q_{\mu}) (T^{\alpha\mu\nu} + g^{\alpha[\mu}q^{\nu]}) - \tfrac{1}{2} T^{\nu}{}_{\mu\sigma} ( T^{\alpha\mu\sigma} + g^{\alpha[\mu}q^{\sigma]})\\ & & + \hat R^{\alpha\nu} - \frac{1}{2}(f^{(\alpha\nu)}+\tfrac{1}{2} f g^{\alpha\nu}) + \di^{[\alpha}q^{\nu]} 
				, \label{eq:a eqn} \\
				\nonumber  b^{\alpha\nu}&= & ({\hat \nabla}_{\mu}-T^{\sigma}{}_{\sigma\mu} + \tfrac{1}{2} q_{\mu})(\theta^{\alpha\mu\nu}-f^{\alpha[\mu}q^{\nu]}) - \tfrac{1}{2} T^{\nu}{}_{\mu\sigma} (\theta^{\alpha\mu\sigma}-f^{\alpha[\mu}q^{\sigma]}) \\ & & -f_{\lambda\mu}(\hat R^{\alpha\lambda\mu\nu} - \frac{1}{2} \phi^{\alpha\lambda\mu\nu}) - f^{\alpha}{}_{\mu} (\di^{[\mu}q^{\nu]}+\frac{1}{2}f^{[\mu\nu]}) - \frac{1}{2} \tau^{\alpha\nu}  , \label{eq:b eqn} \\
				\nonumber   c^{\alpha\beta\nu}&= & ({\hat \nabla}_{\mu}-T^{\sigma}{}_{\sigma\mu})(\hat R^{\alpha\beta\mu\nu} - \frac{1}{2} \phi^{\alpha\beta\mu\nu} ) - \tfrac{1}{2} T^{\nu}{}_{\mu\sigma} ( \hat R^{\alpha\beta\mu\sigma} - \frac{1}{2} \phi^{\alpha\beta\mu\sigma} ) \\ & & + \frac{1}{2} \theta^{[\alpha\beta]\nu} - \frac{1}{2} f^{[\alpha|\mu|} T^{\beta]\nu}{}_{\mu} +  \frac{1}{4} (q^{[\beta} f^{\alpha]\nu} + q_{\mu} g^{\nu[\alpha} f^{\beta]\mu}) , \label{eq:c eqn} \\
				\nonumber d^{\nu}&=& 2 ({\hat \nabla}_{\mu}-T^{\sigma}{}_{\sigma\mu})(\di^{[\mu} q^{\nu]} + \frac{1}{2} f^{[\mu\nu]}) - T^{\nu}{}_{\mu\sigma}(\di^{[\mu} q^{\sigma]} + \frac{1}{2} f^{[\mu\sigma]})  \\ & & + \frac{1}{2}({\hat \nabla}_{\mu}+q_{\mu})(f^{(\mu\nu)}-g^{\mu\nu} f) +\frac{1}{2}f_{\mu\sigma}T^{\sigma\mu\nu} - \frac{1}{2} f_{\mu\sigma} T^{\mu\sigma\nu} + \frac{1}{2}{\hat \nabla}_{\mu} f^{[\mu\nu]} 
				, \label{eq:d eqn}
			\end{eqnarray}
		\end{subequations}
		where the sources in the above are related to those defined in (\ref{eq:Currentj}) by
		\bea
		a^{\alpha\nu} := e_{a}^{\alpha}  a^{a\nu},\\
		b^{\alpha\nu} := e_{{a}}^{\alpha} b^{{a}\nu}, \\
		c^{\alpha\beta\nu}:=e_{{a}}^{\alpha}e_{{b}}^{\beta} c^{{a}{b}\nu}.
		\eea
		Here, $\hat R_{\alpha\beta\gamma\delta}$ is the {\it affine}  Riemann tensor.  The covariant derivative ${\hat \nabla}$ is with respect to the metric compatible connection. 
		Finally we have used the shorthand:
		\begin{align}
			\theta^{\mu\alpha\beta} & := {\hat \nabla}^{\alpha} f^{\mu\beta} - {\hat \nabla}^{\beta} f^{\mu\alpha}, \label{eq:thetadef} \\
			\phi^{\alpha\beta\gamma\delta}  & :=   g^{\gamma[\alpha} f^{\beta]\delta}  - g^{\delta[\alpha} f^{\beta]\gamma}\label{eq:phidef} .
		\end{align}
		Note that at this stage $f^{\mu\beta} = f^{(\mu\beta)} +f^{[\mu\beta]} $ has both a symmetric and anti-symmetric part.

		In the remainder of this paper we refer to the gauge potentials $e^a_\mu,\ell^a_\mu,\omega^{[ab]}_\mu, q_\mu$ as the gravitational Yang-Mills (GYM) fields.
		
			In the case that the torsion vanishes, the right hand sides of the equations of motion (\ref{eq:a eqn}), (\ref{eq:c eqn}) and (\ref{eq:d eqn}) yield an identity that in turn imposes a constraint on the sources on the left, namely:
			$$
			{\hat \nabla}^\alpha a_{\alpha\beta}+c_{\beta\alpha}{}^\alpha-\half d_\beta=0
			\label{eq:SourceIdentity}
			$$

		In the remainder of this paper we restrict the geometry to be torsion-free and the gauge potentials $q_\mu$ and $f_{[\mu\nu]}$ to vanish.\footnote{The general case is considerably more complicated and will be considered elsewhere.}  In this case, (\ref{eq:a eqn}-\ref{eq:d eqn}) reduces to:
		\begin{subequations}
			\bea
			a^{\alpha\nu} & = &  R^{\alpha\nu} - \frac{1}{2}(f^{(\alpha\nu)}+\tfrac{1}{2} f g^{\alpha\nu}), \label{eq:a eqn2} \\
			b^{\alpha\nu} & = & { \nabla}_{\mu}\theta^{\alpha\mu\nu}-f_{\lambda\mu}\left(R^{\alpha\lambda\mu\nu}-\frac{1}{2} \phi^{\alpha\lambda\mu\nu}\right)- 
			\tfrac{1}{2} \tau^{\alpha\nu}, \label{eq:b eqn2} \\
			c^{\alpha\beta\nu} & = &  \nabla_{\mu}\left(R^{\alpha\beta\mu\nu}-\frac{1}{2} \phi^{\alpha\beta\mu\nu}\right)
			+ \frac{1}{2} \theta^{[\alpha\beta]\nu} , \label{eq:c eqn2} \\
			{d^{\nu}} & = & \frac{1}{2} 
			\nabla_{\mu} \left(f^{(\mu\nu)}-g^{\mu\nu} f\right). \label{eq:d eqn2}
			\eea
		\end{subequations}
		In the above, the unhatted covariant derivative operators $\nabla_\mu$ and curvature tensors are those with respect to the Levi-Civita (torsion-free) connection.
		

\section{Conformally coupled scalar}
\label{sec:ConformalScalar}
Our goal is to construct a conformally invariant theory that reduces to the Einstein equations with cosmological constant without the need to add any dimensional coupling constants to the classical action. We therefore {\it conformally} couple a real scalar field $\rho$ to the GYM fields with the following Lagrangian density.  We note that the scalar field  couples only to the metric, or the vierbein, and not to the other GYM connection components.
\bea
L&=&L_{YM}+L_C+L_M;\\
L_C&:=&\half\sqrt{-g}\left[|{\nabla}\rho|^2+\frac{1}{6}\rho^2 {R}-\frac{2\lambda}{4!}\rho^4\right].
\label{eq:LC}
\eea
 $L_M$ is an arbitrary matter Lagrangian that couples only to the metric and not to $\rho$.

We emphasize that ${R}$ is the Levi-Civita curvature scalar with respect to the Levi-Civita connection $ {\Gamma}^{\alpha}{}_{\beta\gamma}:=\half g^{\alpha\delta}\left(g_{\beta\delta,\gamma}+g_{\gamma\delta,\beta}-g_{\beta\gamma,\delta}\right)$.

Under $g_7$ variations of this matter action, one has:
\be
\delta{R}=\nabla^\mu\nabla^\nu\delta g_{\mu\nu}-\Box(g^{\mu\nu}\delta g_{\mu\nu})- R^{\mu\nu}\delta g_{\mu\nu},
\ee
and 
\bea
\delta g_{\mu\nu}&=&\Omega g_{\mu\nu};\nonumber\\
\delta\rho&=&-\half\Omega \rho.
\label{eq:GaugeTransfns}
\eea
We obtain
\be
\delta L_C=\sqrt{-g}
\nabla^\mu\left(\frac{1}{4}\rho^2\partial_\mu\Omega\right) ,
\label{eq:GaugeVariation1}
\ee
{}{where $N_\mu:=\frac{1}{4}\rho^2\partial_\mu\Omega$.}

{We see that the matter action is $g_7$ invariant up to a total divergence}, as expected.  The total divergence term does not affect the equations of motion, which are therefore covariant under the full $g_7$. Note that $N_\mu$ is not the Noether current associated with the dilatation invariance because  there is no physical reason for the total divergence of $N_\mu$ to vanish. 
{As first pointed out in \cite{PhysRevD.91.067501}, the conformally invariant scalar action does not have an associated non-zero Noether current,  the authors of \cite{PhysRevD.91.067501} to label the conformal invariance of this action as ``fake''. 
The gravitational YM action (\ref{eq:action}) on the other hand does  have a non-zero Noether current.

In the present case  the currents $a^{\mu\nu}$, $c^{[\alpha\beta]\nu}$ and $d^\nu$ are zero 
while
\be
b^{\mu\nu}= \half\left[-T^{(\mu\nu)}-\frac{\lambda}{4!}\rho^4 g^{\mu\nu}
+{1/6}(\nabla^\mu\nabla^\nu -g^{\mu\nu}\Box- G^{\mu\nu})\rho^2+M_{\mu\nu}\right],\\
\ee
where $G^{\mu\nu}$ is the Einstein tensor,
\be
T^{\mu\nu}:=\nabla^\mu\rho\nabla^\nu\rho-\half g^{\mu\nu}|\nabla\rho|^2,
\ee
and where $M_{\mu\nu}$ is the stress-energy tensor of `ordinary matter'.  The latter couples only to the metric/vierbein.

The matter equation of motion is:
\be
-\rho\Box\rho -\frac{\lambda}{3!}\rho^4+\frac{1}{6}\rho^2 R=0.
\label{eq:rhoeq2}
\ee

We can solve (\ref{eq:a eqn2}) algebraically for $f_{\mu\nu}$:
\begin{equation}\label{eq:f-eqn}
	\frac{1}{2}	f_{(\alpha\beta)} = 2S_{\alpha\beta},
\end{equation}
where
\begin{equation}
	S_{\alpha\beta} := \tfrac{1}{2}(R_{\alpha\beta} - \tfrac{1}{6} R g_{\alpha\beta}),
\end{equation}
is the Schouten tensor.  The field equations (\ref{eq:c eqn2}) and (\ref{eq:d eqn2}) hold identically on account of the above and the Bianchi identities. 

Using  (\ref{eq:f-eqn}) and the definition (\ref{eq:thetadef}) with the torsion, $q_\mu$ and $f_{[\mu\nu]}$ all set to zero to, the second equation of motion, (\ref{eq:b eqn2}), implies
\begin{eqnarray}\nn
\nabla_{\mu}\theta^{\alpha\mu\nu}&=&2\nabla_\mu\nabla^{[\mu} f^{\nu]\alpha}\\
&=& 16\nabla_\mu\nabla^{[\mu} S^{\nu]\alpha}.
\end{eqnarray}
Furthermore,  (\ref{eq:f-eqn}) and (\ref{eq:phidef}) can be used to show that
\be
R^{\alpha\lambda\mu\nu}-\frac{1}{2}\phi^{\alpha\lambda\mu\nu}=C^{\alpha\lambda\mu\nu},
\ee
where $C^{\alpha\lambda\mu\nu}$ is the Weyl curvature tensor.  Hence the right hand side of (\ref{eq:b eqn2}) is just 
\be
4\left(2\nabla_\mu\nabla^{[\mu} S^{\nu]\alpha}-S_{\lambda\mu}C^{\alpha\lambda\mu\nu}\right),
\ee
which is $4$ times the Bach tensor $B^{\alpha\nu}$.  Thus, the equations of motion boil down to 
\be
g_{YM}^2 b^{\alpha\nu} =4 B^{\alpha\nu},
\ee
or more explicitly and usefully:
\be
\frac{8}{g^2_{YM}}B_{\mu\nu}=-T_{(\mu\nu)}-\frac{\lambda}{4!}\rho^4 g_{\mu\nu}+M_{\mu\nu}
+\frac{1}{6}(\nabla_\mu\nabla_\nu -g_{\mu\nu}\Box- G_{\mu\nu})\rho^2.
\label{eq:EOMbach}
\ee

Since the Bach tensor is identically traceless, the trace of the right hand side of (\ref{eq:EOMbach}) must vanish on shell, i.e.
\be
-\rho\Box\rho+\frac{1}{6}\rho^2  R-\frac{\lambda}{3!}\rho^4 +M=0,
\ee
where $M:=g^{\mu\nu} M_{\mu\nu}$.  
Thus consistency with the $\rho$ equation of motion requires $M=0$.  

When $\rho(x)=\rho_0=$ constant the scalar equation of motion reduces to 
\be
E_\rho=\frac{\rho_0}{6}(R+\lambda\rho^2_0)=0.
\ee
or, if $\rho\neq0$:
\be
R=\lambda\rho^2_0.
\ee


Different gauge choices for the scalar correspond to different frames for the metric, with $\rho=\rho_0$ corresponding to the so-called Einstein frame. As long as further matter couplings preserve the conformal invariance, the physics must be the same in all frames. However for matter that is not conformally invariant, the choice of frame must be made on physical or observational grounds.

 
We now discuss the physical significance of the dimensionless coupling constants $g^2_{YM}$ and $\lambda$.
Dividing both sides of (\ref{eq:EOMbach}) by $\rho_0^2$ and rearranging terms we get:
\bea
G^{\mu\nu} +\frac{\lambda}{4}\rho_0^2g^{\mu\nu}+\frac{48}{ g^2_{YM}\rho^2_0 }B^{\mu\nu}&=&\frac{6}{\rho_0^2}M^{\mu\nu}.\\
\label{eq:EOMbach3}
\eea

Assuming the phenomenological stress tensor  $M_{\mu\nu}$ represents ordinary matter, one sees that Newton's constant is given by:
\be
8\pi G = \frac{6}{\rho_0^2},
\label{eq:PlanckScale}
\ee 
with corresponding Planck length
\be
l_{pl} = \sqrt{G} =\sqrt{\frac{6}{8\pi\rho_0^2}} 
\label{eq:PlanckScale2}
\ee
The cosmological constant is:
\be
\Lambda = \frac{\lambda}{4}\rho^2_0 =\frac{3}{16\pi}\frac{\lambda}{ l_{pl}^2}.
\label{eq:CosmologicalConstant}
\ee
Since $B^{\mu\nu}$ is fourth order in derivatives, and the Einstein tensor is second order, one can heuristically argue that the Bach contribution is negligible when the characteristic momentum scale $k$   of the metric fluctuations satisfy:
\bea
&&\frac{48}{  g^2_{YM}\rho^2_0 }k^4<<k^2;
\\
&&k^2<<k^2_B\equiv\frac{g^2_{YM}\rho^2_0 }{48}=\frac{ g^2_{YM}}{64\pi}\frac{1}{\ell_{pl}^2}.
\label{eq:BachFadeOut}
\eea

To recap, the conformally invariant equations of motion are:
\bea
&&-\rho\hat\Box\rho -\frac{\lambda}{3!}\rho^4+\frac{1}{6}\rho^2\hat R=0,\nonumber\\
&&\frac{8}{g^2_{YM}}B^{\alpha\nu}=-T^{(\mu\nu)}+M^{\mu\nu}-\frac{\lambda}{4!}\rho^4 g^{\mu\nu}
+\frac{1}{6}(\hat\nabla^\mu\hat\nabla^\nu -g^{\mu\nu}\hat\Box- G^{\mu\nu})\rho^2
\label{eq:EOMbachConf}
\eea
{When $M^{\mu\nu}$ is traceless one can always do a conformal transformation to the Einstein frame in which $\rho=\rho_0=$constant so that the theory is gauge equivalent to Bach plus Einstein with cosmological constant. }

The vacuum expectation value, $\rho_0$, of the scalar determines Newton's constant and hence the Planck scale, while the dimensionless coupling constant, $\lambda$, determines the cosmological constant in Planck units. Finally, the dimensionless Yang-Mills coupling constant then determines the momentum scale $k_B$ at which the Bach tensor becomes important. We have therefore succeeded in our goal of constructing a theory in which the dimensional constants in Einstein's theory are obtained from a scale invariant ``gravitational'' action via symmetry breaking.  We note here that Bars et. al. (\cite{Bars_2014}) constructed a model with similar outcomes for emergent gravity, but with less generality, i.e. with no coupling to the panoply of possible geometrical non-Einsteinian fields.  \\

\section{Cosmological solution}
\label{eq:CosmologicalSolution}
\subsection{Metric Ansatz}
We focus on the sector of the theory in which the torsion, $q_\alpha$, and  $f_{[\alpha\beta]}$ all vanish and consider the metric {\it ansatz}
\bea
ds^2&=&-dt^2 +a^2(t)\left[\frac{dr^2}{f(r)}+r^2(d\theta^2+\sin^2{\theta} d\phi^2)\right].\label{eq:modfrw}
\eea
The Einstein tensor, Bach tensor and Kretzschmann scalar for this metric are given in the Appendix \ref{sec:Geometry}.

In Einstein gravity the function $f(r)=1-k r^2$ and the metric is one of the FRW spacetimes, {where the curvature of the spatial metric is zero for $k=0$, positive for $k>0$ and negative for $k<0$.} The function $a(t)$ is the scale factor. This is also a solution with vanishing Bach tensor in the present theory,
but there exists a more general class of spherically symmetric inhomogeneous solutions whose Bach tensor is non-zero.

\subsection{Spherically symmetric anisotropic perfect fluid source}
In addition to the conformally coupled scalar field $\rho(t,r)$, as per the previous section, we consider an anisotropic fluid with the following phenomenological stress-energy tensor:
\be
M^{\mu\nu}=-\left(({p_\perp}(t,r)-{p_\|}(t,r))V^\mu V^\nu+(\epsilon(t,r)+{p_\|}(t,r))U^\mu U^\nu+{p_\|}(t,r)g^{\mu\nu}\right),\label{eq:Matter}
\ee
where $\epsilon(t,r)$ is the fluid energy density, {There are two equal transverse pressures, ${p_\|}(t,r)$,  that are  tangent to the two-sphere at fixed areal radius $r$. 
They must be equal to respect spherical symmetry.  $p_\perp(t,r)$ is the radial or longitudinal pressure and is perpendicular to the Killing two sphere.}  When ${p_\perp}(t,r)={p_\|}(t,r)$ the fluid is isotropic.

In spherical coordinates $(t,r,\theta,\phi)$:
\be
M^{\mu}{}_{\nu} = \hbox{diag}(-\epsilon(t,r), +{p_\perp}(t,r), +{p_\|}(t,r), +{p_\|}(t,r)).
\ee

\subsection{Metric components}
In the equations of motion (\ref{eq:EOMbachConf}), the stress-energy tensor $M^{\mu\nu}$ is given by (\ref{eq:Matter}).  
There is one off-diagonal equation of motion, namely
\be
8 B^{tr}=-T^{(tr)}+M^{tr}-\frac{\lambda}{4!}\rho^4 g^{tr}
-\frac{1}{6}(\hat\nabla^t\hat\nabla^r + G^{tr})\rho^2,
\ee
which for the metric {\it ansatz} (\ref{eq:modfrw}) reduces to:
\be
\frac{1}{a^2(t)}\left[2\rho\rho_{,t,r}a^2(t)-4\rho_{,t}\rho_{,r}a^2(t)-2 a(t) \rho\rho_{,r} a_{,t}\right]=0.
\ee

This has three solutions with $a(t)\neq 0$:
\begin{enumerate}[label=(\roman*)]
	\item $\rho(t,r)=0$;
	\item $\partial_r\rho=0$, but $\rho(t)$ is arbitrary;
	\item $\rho(t,r)=A(r)/a(t)$.  
\end{enumerate}
{}{The trivial solution {\it (i)} corresponds to Weyl-Bach gravity.}
In case (iii), $a(t)$ is not determined by $E_\rho=0$.  It remains
arbitrary.  If we consider the simple case where $A(r)=$  constant,
then it  turns out that for the cosmological setting, the fluid has
negative energy for all values of the parameters.  Furthermore, the
Einstein frame could only be realized when both $A(r)$ {\it and}
$a(t)$ are constant. That is, the geometry is static. 
In the following we therefore only consider case (ii).

The equation of motion obtained by variation of $\rho$ is:
\be
-\rho\Box\rho -\frac{\lambda}{6}\rho^4+\frac{1}{6}\rho^2 R=0.
\ee
For case (ii) it reduces to:
 \be
\frac{\rho}{a^2}\left(a^2\ddot\rho+\rho (\dot a^2 +a \ddot a)+3\dot\rho a \dot a-\frac{\lambda}{6}\rho^3 a^2+\frac{\rho}{3 r^2}(-r f'-f +1)\right)=0.
\label{eq:ScalarEoM2}
\ee
Derivatives with respect to the coordinates $t,r$ are denoted, respectively, by an over dot and a prime.  One can  separate the variables $(t,r)$ in the above to get the condition:
\be
-r f'-f+1 =3 k r^2,
\ee
where k is a separation constant and the factor of 3 is inserted for convenience.   This has the solution
\be
f(r)=1-k r^2 +\frac{Q}{r},
\ee
where $Q$ is a constant of integration.  We see that in this case the spacetime geometry deviates from FRW by the presence of the term $Q/r$.  If $Q=0$, then the Bach tensor vanishes (see Eq.~(\ref{eq:Kr})). The metric (\ref{eq:modfrw}), with $f(r)$ as above, has appeared in the literature, in the context of anisotropic and inhomgeneous cosmology.  See, for example Bayin (\cite{Bayin1986}).

When $\rho(t)\neq 0$ the scalar field equation of motion (\ref{eq:ScalarEoM2}) now reduces to an ordinary differential equation in $t$:
\be
E_\rho:=a^2\ddot\rho+\rho (\dot a^2 +a \ddot a)+3\dot\rho a \dot a-\frac{\lambda}{6}\rho^3 a^2+k\rho=0.
\label{eq:rho}
\ee

The equations of motion $E_{\mu\nu}=0$ obtained by varying the metric components are:
\bea
E_{tt}=0\implies\qquad  \epsilon &=&\frac{1}{24 a^4}\left[{N(t)} +72\frac{Q^2}{g_{YM}^2 }r^{-6}\right];\label{eq:Ett}\\
E_{rr}=0\implies\qquad  {p_\perp} &=&\frac{1}{24 f(r) a^4}\left[{M(t)}+\frac{4 Q}{r^3}\left(\rho^2 a^2 +\frac{24 k}{ g_{YM}^2} \right) -\frac{24}{r^6}\frac{Q^2}{ g_{YM}^2 }\right];\label{eq:Err}\\
E_{\theta\theta}=0\implies\qquad {p_\|} &=&\frac{g_{YM}^2 r^2}{24  a^4}\left[{M(t)}-\frac{2 Q}{r^3}\left(\rho^2 a^2 +\frac{24 k}{ g_{YM}^2} \right) +\frac{48}{r^6}\frac{Q^2}{ g_{YM}^2 }\right],\label{eq:Ethetatheta}
\eea
where we have put  $M_{\mu\nu}$ on the left hand side of the equations above, so the right hand side contains the ``geometrical'' contributions from the Bach and Einstein tensors as well as the conformal scalar. The Bach terms are those quadratic in $Q$. We have also defined
\bea
N(t)&:=&-\lambda\rho^4 a^4+12\rho^2(a\dot{a})^2 +12 k {\rho^2}{a^2} +24\rho\dot{\rho}{a^2}{a\dot{a}} +12\dot{\rho}^2 a^4;
\label{eq:N(t)}\\
M(t)&:=&-\lambda\rho^4 a^4-4\rho^2(a\dot{a})^2 +4 k {\rho^2}{a^2} +16\rho\dot{\rho}{a^2}{a\dot{a}} -4\dot{\rho}^2 a^4+8\rho^2 a^2 ( a \ddot{a} + (\dot a)^2 )+8\rho\ddot{\rho} a^4.\nonumber\\
\label{eq:M(t)}
\eea
Eqs.(\ref{eq:N(t)}), (\ref{eq:M(t)})  and (\ref{eq:rho})  imply that
\be
\frac{1}{24 a^4}(-N(t)+3M(t))=g_{YM}^2 E_\rho.
\label{eq:OnShellTrace}
\ee

Calculating  the trace of the equations of motion, $E:=g^{\mu\nu}E_{\mu\nu}$  from Equations (\ref{eq:Ett},\ref{eq:Err},\ref{eq:Ethetatheta}) yields
\be
E=\frac{g^2_{YM}}{24 a^4}\left[-N(t)+3M(t)\right]+g^2_{YM}\left[\epsilon - 2{p_\|} - {p_\perp}\right]=0.
\ee
Using (\ref{eq:OnShellTrace}) we conclude that $M_{\mu\nu}$ must be traceless on-shell, namely
\be
\epsilon - 2{p_\|} - {p_\perp}=0.
\label{eq:MatterTrace}
\ee
If the phenomenological stress-energy tensor were derived from a conformally invariant action this condition would be identically satisfied on shell. In the present case (\ref{eq:MatterTrace}) is an extra condition on the phenomenological stress-energy tensor. It imposes an equation of state that requires the matter to be in effect radiation-like.

We now calculate $\dot N(t)$ from Eq.~(\ref{eq:N(t)})  to get
\be
\dot{N}(t)=4\left(a\dot{a}+a^2\frac{\dot\rho}{\rho}\right)E_\rho.
\ee
Hence, on-shell, we find that 
\bea
N(t) = 3 M(t) = 24 N,
\label{eq:N3M}
\eea
where $N$ is a constant and  the numerical factor of $24$ is included for convenience.

To recap, we have  solved for the metric functions $f(r)$ and $a(t)$ in terms of $\rho(t)$ using only the scalar equation, revealing two independent constants of motion: the separation constant $k$ and the constant 
\bea
N=3M.
\label{eq:N3M2}
\eea
 In addition, we have shown that the equations of motion require the phenomonological stress-energy tensor to be traceless, as required by the conformal invariance of the vacuum theory.

The equations of motion for (\ref{eq:Ett}, \ref{eq:Ethetatheta},\ref{eq:Err}) for $\epsilon(t,r),p_\perp(t,r), p_\|(t,r)$ yield respectively:
\bea
\epsilon(t,r)&=&\frac{1}{a^4}\left(N+\frac{3 Q^2}{ g_{YM}^2}r^{-6}\right).\label{eq:epsoleom};\\
p_\|(t,r)&=&\frac{1}{ a^4}
\left[\frac{N}{3}+\frac{ Q}{12 r^3}\left(\rho^2 a^2+24 \frac{k}{g_{YM}^2}\right)-\frac{2}{r^6}\frac{Q^2}{ g_{YM}^2}\right];\label{eq:p1soleoma}\\
p_\perp(t,r)&=&\frac{1}{a^4}
\left[\frac{N}{3}-\frac{ Q}{6r^3}\left(\rho^2 a^2+24 \frac{k}{g_{YM}^2}\right)+\frac{1}{r^6}\frac{Q^2}{ g_{YM}^2}\right].\label{eq:p2soleomb}
\eea
One can verify that these expressions satisfy the traceless condition
\be
M^\mu{}_\mu(t) = \epsilon(t) - 2 p_\|(t) - p_\perp(t)
=0.
\label{eq:TraceCondition}
\ee
As well, the $a^{-4}$ dependence in the leading terms (large $r$ behaviour)  in both ${p_\|}(t,r)$ and ${p_\perp}(t,r)$ is expected from traceless (radiation-like) matter.

The above analysis shows that for our choice of phenomenological energy momentum tensor and metric tensor, the energy density and both pressures are uniquely determined in terms of the scale factor $a(t)$ and conformal scalar $\rho(t)$ by the equations of motion equations or equivalently by the matter energy momentum conservation laws and trace condition. The trace condition is an extra constraint that is not present in standard $\Lambda$FRW cosmology. It requires the equation of state of the matter we have introduced to be that of radiation.

The metric components on the other hand are only determined up to a conformal transformation of the metric. One can fix either $\rho(t)$, in which case $a(t)$ is determined, or one can fix $a(t)$ in which case $\rho(t)$ is determined. As suggested above, the simplest and most natural choice is the so-called Einstein frame $\rho = \rho_0=$constant. As verified explicitly in Section \ref{sec:GeneralRho} other solutions are related by a conformal transformation of the scalar and the metric. 

The energy density, longitudinal and transverse pressures transform as $L^{-2}$ under a dilatation due to the overall factor of $a^4$,  but are otherwise form invariant. Note that the term $\rho^2 a^2$ is invariant under conformal transformations

\subsection{  Einstein Frame Solution and Physical Interpretation}

We saw  in section(\ref{sec:ConformalScalar}) that with $\rho=\rho_0$, the theory approximates Einstein gravity the {Einstein frame}.
Equations { (\ref{eq:N(t)}) and (\ref{eq:M(t)})  can be rewritten to correspond to the standard first and second Friedmann equations, respectively: 
	\bea
	\frac{{\dot{a}}^2}{a^2} &:=&\frac{N}{12 a^4\rho_0^2}+\frac{\lambda\rho_0^2}{12}- \frac{k}{a^2};
	\label{eq:Friedman1c}
	\eea
	\bea
	\frac{\ddot a}{a}&=&  
	- \frac{(N+3M)}{24 a^4\rho_0^2}+\frac{\lambda\rho_0^2}{12}.\nonumber\\
	&=&  
	- \frac{N}{6 a^4\rho_0^2}+\frac{\lambda\rho_0^2}{12}.
	\label{eq:Friedman2b}
	\eea
	These equations are of the form of the standard homogeneous, isotropic Friedmann equations (see for example,\cite{Carroll:2004st}) for the scale factor $a(t)$ with cosmological constant $\Lambda=\frac{\lambda\rho_0^2}{4}$,  three curvature $k$ and radiation type stress tensor with $\rho_{rad} \sim N/a^4$ and $p_{rad} \sim M/a^4 = \rho_{rad}/3$ .

	We now look at the solutions of (\ref{eq:Friedman1c}) and (\ref{eq:Friedman2b}).
	In our cosmological setting, we find from $E_\rho=0$ that 
	\be
	a^2(t)=C_1\exp{(\alpha t)}+C_2\exp{(-\alpha t)}+\frac{2k}{\alpha^2}.
	\ee
	If we are to have the scale factor $a^2(t)$ non-negative, the integration constants $C_1, C_2$ must both be positive, so that without loss of generality, 
	\be
	a^2(t)=C\cosh{(\alpha (t-t_0))}+\frac{2 k}{\alpha^2},\label{eq:asq}
	\ee
	where $\alpha^2:=\rho_0^2\lambda/3=\tfrac{4}{3}\Lambda$, $C_1,C_2, C, t_0$ are integration constants. We henceforth set $t_0=0$ without loss of generality.
	
	The relationship between the integration constants $C$ in (\ref{eq:asq}) and $N$ is:
	\bea
	N&=& 36k^2-C^2\lambda^2\rho_0^4\\
	&=& 36k^2 - 16 C^2 \Lambda^2.
	\label{eq:NC}
	\eea

}

The resulting spacetime has constant Ricci scalar $R=\frac{4}{3}\Lambda$. The components of the Einstein tensor and Bach tensor are given in Appendix \ref{sec:Geometry}. The solutions are not Einstein spaces, since the latter  have zero Bach tensor. There is a curvature singularity at $r=0$ (see the Kretschmann scalar in Eq.~(\ref{eq:Kr})), but in the large $r$ limit, the solution reduces to the $\Lambda$FRW cosmology. {We see that there is no big bang singularity here, as long as $k\geq 0$.}\\


In the Einstein frame, we have
\bea
\epsilon(t,r)&=&\frac{1}{24 {}\lambda a^4(t)}\left(N+72 \frac{Q^2}{r^6}\frac{\lambda}{g_{YM}^2}\right);\\
{p_\|}(t,r)&=&\frac{1}{72  {}\lambda a^4(t)}\left[N-6 \frac{Q\lambda\rho_0^2}{r^3}\left(a^2(t)+24 \frac{k}{\rho_0^2}\right)+144  \frac{Q^2}{r^6}\frac{\lambda}{g_{YM}^2}\right];\\
{p_\perp}(t,r)&=&\frac{1}{72  {}\lambda a^4(t)}\left[N+12 \frac{Q\lambda\rho_0^2}{r^3}\left(a^2(t)+24 \frac{k}{\rho_0^2}\right)-72{} \frac{Q^2}{r^6}\frac{\lambda}{g_{YM}^2}\right].
\eea
The traceless condition is satisfied for all parameter values.
As expected, for large $r$ $M_{\mu\nu}$ corresponds to the stress energy tensor of a homogeneous and isotropic radiation dominated spacetime.

We also see from 
(\ref{eq:NC}) that  $N<0$ and hence the energy density goes negative for large $r$  unless: 
\be
k>\frac{C \lambda\rho_0^2}{6}.
\label{eq:kC condition}
\ee

The energy density is positive for all $r$, as long as (\ref{eq:kC condition}) is satisfied. Violations of the energy conditions can and do occur for small $r$, specifically when the $Q$ dependent terms dominate the energy and pressures. This is to be expected in an effective theory such as this because at small distances one expects quantum corrections to become significant. These quantum effects can in principle violate any and all of the energy conditions \cite{Visser2000}. A better understanding of the physical significance of these violations in the context of the present model requires the construction of a more physical cosmological model than the illustrative one studied here.

We stress again that there are no dimensional parameters in the action gravitational action. The cosmological constant derives from the values of $\lambda$ and  $\rho_0$ in the solution. The only scale invariant solution has $\rho_0=0$ which is unphysical, not least because the associated Newton constant is zero. There is a continuum of non-scale invariant solutions so that the cosmological constant is in essence a consequence of spontaneous symmetry breaking, even though the mechanism is somewhat different from the standard picture that occurs in, say, scalar QED.

\begin{figure}[h]
	\centering
	\includegraphics[width=0.5\textwidth]{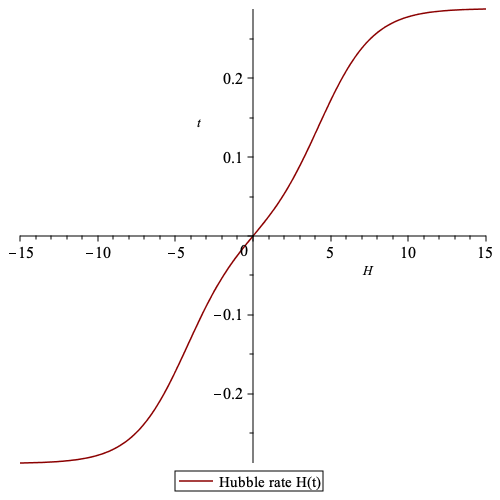}
	\caption{Graph of $H(t)$
	 The solution parameters are: 
	$k=1,Q=-1,C=1,t_0=0,\rho_0=1$. In terms of physical parameters: $G=l^2_{pl}=\tfrac{3}{4\pi}$, $\Lambda = \tfrac{1}{4}$, $r_{Bach} =\tfrac{1}{k_B}=\sqrt{48}$
		\label{fig:Figure 1}}
\end{figure}
\begin{figure}[h]
	\centering
	\includegraphics[width=0.5\textwidth]{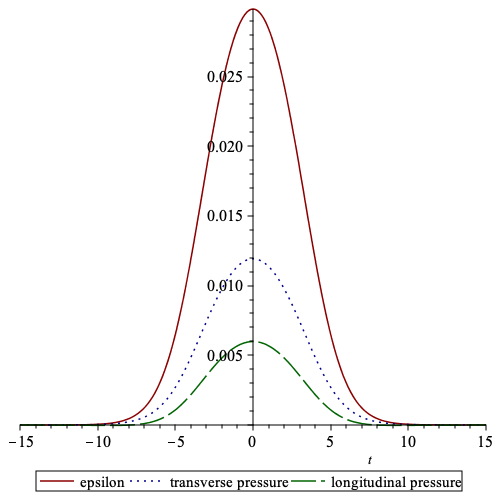}
	\caption{Plot of energy density and pressures in Einstein frame, with parameter values $g_{YM}=1,Q=-1,r=3,C=1,t_0=0,k=1,\omega=1,\rho_0=1$. {}{In terms of physical parameters: $G=l^2_{pl}=\tfrac{3}{4\pi}$, $\Lambda = \tfrac{1}{4\pi}$, $r^2_{Bach} = \tfrac{6}{\pi} < r^2 =9$.The curves $\epsilon, p_\perp, p_\|$, respectively, are plotted in red, green, blue.}
		\label{fig:Figure 2}}
	\end{figure}
\begin{figure}[h]
	\centering
	\includegraphics[width=0.5\textwidth]{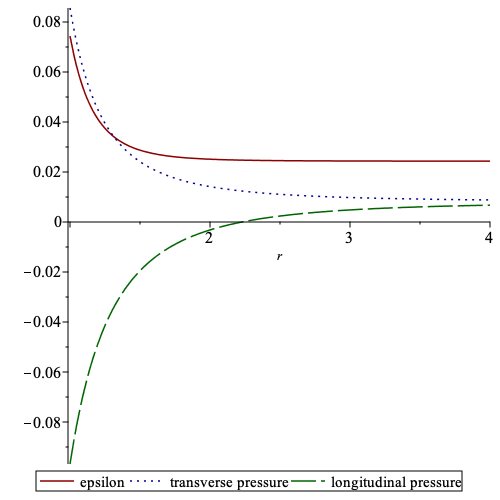}
	\caption{Plot of energy density and pressures versus $r$ in Einstein frame, with parameter values $g_{YM}=1,Q=-1,t=2,C=1,t_0=0,k=1,\omega=1,\rho_0=1$. {In terms of physical parameters: $G=l^2_{pl}=\tfrac{3}{4\pi}$, $\Lambda = \tfrac{1}{4\pi}$, $r^2_{Bach} = \tfrac{6}{\pi} $.The curves $\epsilon, p_\perp, p_\|$, respectively, are plotted in red, green, blue.}\label{fig:Figure 3}}
	\end{figure}



\section{Discussion}
\label{sec:Discussion}

We have {}{extended a previous analysis of a gauge theory of gravity in which the action is quadratic in the Yang-Mills curvature, with conformal group SO(4,2) as the gauge group.} In the usual YM gauge theories, the spacetime metric is a given fixed background. {}{Instead of being fixed, the metric in this theory} is made dynamical by identifying it with the fields gauged by the translation generator. This breaks the full SO(4,2) gauge invariance  to that generated by the Lorentz rotations, special conformal transformations and dilatations.  Under certain restrictions, the vacuum equations of motion are solved by the solutions of the vacuum equations of Weyl squared gravity.  

{}{Our goal here was to extend this model so as to} generate all dimensional constants, including Newton's constant and the cosmological constant via symmetry-breaking from a second order, scale invariant theory. This goal was realized by adding to the Yang-Mills-type action a conformally coupled scalar \cite{Bars_2014}.
In the Einstein frame, the resulting theory resembles Weyl-Einstein gravity with non-zero cosmological constant {}{and contains Einstein gravity in the long wavelength limit.} The values of the Newtonian constant/Planck scale, the cosmological constant and the scale at which higher derivative terms contribute to the equations of motion are all determined by the three parameters in the theory: the Yang-Mills coupling constant, the scalar self-coupling and the vacuum expectation value of the scalar. The latter is the only length scale in the theory and comes about via the breaking of the conformal invariance of the model. Given the existence of three parameters for the three physical constants, it is possible to make the theory match the observed values of the Planck constant and cosmological constant while ensuring that the Weyl terms do not affect known physics.

We then added a Weyl invariant phenomenological stress tensor which yielded via an appropriate metric ansatz a spherically symmetric inhomogeneous cosmological solution with non-trivial Bach tensor. The most general solution has a curvature singularity at the origin of the spherical symmetry, but approaches the standard $\Lambda$FRW at large distances. In the Einstein frame, the cosmological Big Bang is replaced by a Big Bounce, whose minimum radius is determined by the scalar self-coupling and the vacuum expectation value of the scalar field.

Our results raise many interesting questions, including the following:
\begin{itemize}
 \item In order to fully understand the dynamical consequences of the symmetry breaking it is necessary to find solutions that include the other modes in the gravitational action that have here been set to zero. It is significant that any conserved charge associated with scale invariance must come from these extra modes.
\item Although spherically symmetric geometries have been examined in conformal gravity, we suspect that interesting physics will arise  in our model when the other so(4,2) gauge potentials and conformally invariant matter are included.  It is therefore of interest to explore the questions of uniqueness and asymptotic behaviour of static spherically symmetric solutions, i.e. black holes candidates with non-zero Bach tensor.
\item Can one construct a more realistic model containing other forms of  interesting conformally invariant matter, such as so(4,2) Higgs multiplets, that couple to all the so(4,2) gauge potentials, and not just the metric as in the present case?
\item Our model has a scale invariant action with no dimensional coupling constants, and leads to second order equations of motion before any restrictions are applied. This suggests, but by know means guarantees, that the model might in fact be perturbatively renormalizable and free of ghosts. 
 A study of the full linearized spectrum and perturbative quantization is required to settle this important issue.
 \end{itemize}
}
The results presented here suggest that these and other issues are worthy of continued investigation.

\begin{acknowledgments}
	
	GK gratefully acknowledge that this research was
supported in part by Discovery Grant number 2018-0409 from the Natural Sciences and Engineering Research Council of Canada.
	
\end{acknowledgments}  

\appendix
\section{Geometry}
\label{sec:Geometry}

In this Appendix we display the relevant curvature tensors and scalars for the cosmological metric (\ref{eq:modfrw}).
The Ricci scalar is
\be
R=6\left(\frac{\ddot a}{a}+\frac{\dot a^2}{a^2} +\frac{ k}{a^2}\right).
\ee

The Einstein tensor $G_{\mu\nu}$ has non-zero components:
\bea
G_{tt}&=&{3}\left(\frac{\dot a^2}{a^2}+\frac{k}{a^2}\right);  \\
G_{rr}&=& \frac{1}{ f(r)}\left[- ( 2 a \ddot a+ \dot a^2+k ) +\frac{Q}{r^3})\right] ; \\
G_{\theta\theta}&=& r^2\left[- ( 2 a \ddot a+ \dot a^2+k ) -\frac{Q}{r^3})\right] ; \\
G_{\phi\phi}&=&\sin^2{\theta}G_{\theta\theta} ,\\
\eea
where $f(r):=1-k r^2+\frac{Q}{r}$.

The Bach tensor is:
\bea
B^{\mu\nu} =
\left[
\begin{array}{cccc}\frac{3~Q ^{2}}{8~a^{4}~r ^{6}} & 0 & 0 & 0 \\0 & -\frac{\left(-k ~r ^{3}+Q +r \right)~Q ~\left(-4~k ~r ^{3}+Q \right)}{8~a^{6}~r ^{7}} & 0 & 0 \\0 & 0 & \frac{Q ~\left(-k ~r ^{3}+Q \right)}{4~a^{6}~r ^{8}} & 0 \\0 & 0 & 0 & \frac{Q ~\left(-k ~r ^{3}+Q \right)}{4~a^{6}~r ^{8}~\sin \left(\theta \right)^{2}} \\
\end{array}\right]
\eea

Finally, the Kretschmann scalar is
\be
Kr:=R^{\mu\nu\alpha\beta}R_{\mu\nu\alpha\beta}=K_0(t)+\frac{6Q^2}{a^4 r^6}.
\label{eq:Kr}
\ee
where
\be
K_0(t):={12}\left[(\dot H+H^2)^2+H^4+\left(\frac{k}{a^2} +H^2\right)^2\right],
\ee
where $H:=\dot a/ a$.  

\section{General Frame Solution of $E_\rho=0$}
\label{sec:GeneralRho}
In the equation $E_\rho=0$, write $\rho(t):=\rho_0\dot{b}(t)$, where $b(t)$ is arbitrary.  
We consider only $\lambda>0$.\footnote{For $\lambda<0$, the scale factor is oscillatory.}

Then according to Maple, the general solution for $a(t)$ which has Lorentzian signature is:
\be
a^2(t)=\frac{1}{\dot{b}^2}\left[C\cosh{(\alpha b(t))}+\frac{2 k}{\alpha^2}\right],
\label{eq:vGeneral}
\ee
with $\alpha:=\rho_0\sqrt{\frac{\lambda}/{3}}$.  The metric is of course given by
\be
ds^2=\left[-d{t}^2+a^2( t)\left(\frac{dr^2}{1-kr^2+Q/r}+r^2(d\theta^2+\sin^2{\theta}d\phi^2)\right)\right].
\label{eq:dsGeneral}
\ee
 Note that this does not involve a conformal transformation- it is just the solution of the equation of motion $E_\rho=0$ for {\it arbitrary} $\rho(t)$. Now define a new time-coordinate 
\be
\bar{t}:=b(t),
\ee
so that 
\be
\frac{d\bar t}{dt}=\dot{b}(t).
\ee
Then we find
\be
a^2(\bar t)=\frac{1}{\left(\frac{d\bar t}{dt}\right)^2}\left[C\cosh{(\alpha \bar t)}+\frac{2 k}{\alpha^2}\right],
\ee
In terms of the new coordinate $\bar t$, the metric is given by:
\be
d\bar{s}^2=\left(\frac{1}{\dot b}\right)^2\left[-d\bar{t}^2+{\bar a^2}(\bar t)\left(\frac{dr^2}{1-kr^2+Q/r}+r^2(d\theta^2+\sin^2{\theta}d\phi^2)\right)\right],
\label{eq:dsBar}
\ee
where
\be
\bar a^2(\bar t)=\left[C\cosh{(\alpha \bar t)}+\frac{2 k}{\alpha^2}\right]
\label{eq:vBar}
\ee

We recover the Einstein frame by choosing $b(t)=t$, so that $\rho(t)=\rho_0$.

By comparing (\ref{eq:dsBar}) and (\ref{eq:vBar}) to (\ref{eq:modfrw}) and (\ref{eq:asq}) we see that, as argued above, the general solution (\ref{eq:vGeneral}, \ref{eq:dsGeneral}), when written in terms of the new time $\bar t$, is conformal to the solution with $\rho(t)=\rho_0$ with conformal factor $e^{\Omega(t) }= (\dot{b}(t))^{-2}$. We see from the above that this conformal transformation provides the Jacobian of the transformation between the $t$ and $\bar t$ coordinates. We expect that as long as this transformation is non-singular, the qualitative behaviours of the solution (i.e. bounce and asymptotic expansion) will be preserved.\\

\bibliography{gravity_yang_mills}

\end{document}